\documentclass[preprint2]{aastex}

\shorttitle{MASSIVE STAR FORMATION IN THE 3-KPC ARMS}
\shortauthors{Green et al}

\begin{document}

\title{HIGH-MASS STAR FORMATION IN THE NEAR AND FAR 3-KPC ARMS}

\author{J. A. Green$^{1}$, N. M. McClure-Griffiths$^{1}$, J. L. Caswell$^{1}$,\\
S. P. Ellingsen$^{3}$, G. A. Fuller$^{2}$, L. Quinn$^{2}$ and M. A. Voronkov$^{1}$}
\affil{$^{1}$Australia Telescope National Facility, CSIRO, PO Box 76, Epping, NSW 1710, Australia;\\
$^{2}$Jodrell Bank Centre for Astrophysics, Alan Turing Building, University of Manchester, Manchester, M13 9PL, UK;\\
$^{3}$ School of Mathematics and Physics, University of Tasmania, Private
Bag 37, Hobart, TAS 7001, Australia\\}

\begin{abstract}
We report on the presence of 6.7-GHz methanol masers, known tracers 
of high-mass star formation, in the 3-kpc arms of the inner Galaxy. We 
present 49 detections from the Methanol Multibeam Survey, the largest 
Galactic plane survey for 6.7-GHz methanol masers, which coincide in 
longitude, latitude and velocity with the recently 
discovered far-side 3-kpc arm and the well known near-side 3-kpc arm. 
The presence of these masers is significant evidence for 
high-mass star formation actively occurring in both 3-kpc arms.
\end{abstract}



\section{Introduction}
The near 3-kpc arm \citep{vanwoerden57}, existing within $15^{\circ}$ of the Galactic centre,  was long believed to be devoid of 
any significant star formation \citep{lockman80}. The recent discovery of a far 3-kpc 
arm within the same range of longitudes \citep{dame08} prompted further thought to this end, the authors speculating 
the narrow thickness of CO and HI in the two arms is evidence that the 3-kpc arms lack 
significant levels of star formation.

However, from the theoretical viewpoint of density waves \citep{lin64, roberts69}, we 
would expect the leading edges of any spiral arms, regardless of location, to foster 
some level of star formation. Furthermore, although primarily occurring in the spiral 
arms, star formation can exist outside of them. It is only the formation of high-mass stars, 
with their short lifetimes, which will be restricted to the spiral arms themselves (and a Galactic bar).

Observational evidence already exists for a few regions of star formation within the near 3-kpc arm.
\citet{caswell87b} detected approximately five H{\sc ii} regions associated with the near arm and 
\citet{cersosimo90} inferred the presence of more diffuse H{\sc ii} regions through H166$\alpha$ recombination line detections.
Additionally, \citet{busfield06} kinematically located a group of infrared colour selected massive 
young stellar objects from the Red MSX Source survey with the near arm. These few previous observations are restricted to only the near arm. To fully address the question of star formation in both 3-kpc arms, a reliable and readily observable tracer of high-mass star formation is required. 

The methanol maser transition at 6.668-GHz has, since its discovery by \citet{menten91}, been demonstrated to be exclusively associated with high-mass star-formation regions \citep{minier03, xu08}. 6.7-GHz methanol masers are typically observed towards the early hot core phases of the star formation process \citep[e.g.][]{minier05,hill05,purcell06}, and have been found in association with other tracers of the early stages of high-mass star formation, such as infrared dark clouds \citep{ellingsen06} and Extended Green Objects \citep{cyga08}.
They are also found in association with ultra-compact regions of ionized hydrogen and are believed to be pumped by infrared radiation  \citep{cragg05}. This species of maser is bright, widespread and has already been shown to be present towards the Galactic centre, possibly within the inner Galactic bar \citep{caswell96c}, demonstrating the exotic environment of this region is not suppressing high-mass star formation. 

Individual 6.7-GHz methanol maser sites tend to exhibit emission features over a range of velocities. 
The velocities of the peaks of the emission features are all typically within a few km\,s$^{-1}$ (to at most $\sim$10\,km\,s$^{-1}$)  of the velocity of the peak of CS (2-1) emission, a reliable tracer of dense gas \citep[e.g.][]{walt07}. 6.7-GHz methanol masers therefore have a good velocity correlation with the molecular clouds in which they live and should trace the same regions in longitude-velocity space as the molecular emission attributed to the near and far 3-kpc arms in  \citet{dame08}.

\section{The existence of high-mass star formation in the 3-kpc arms}
\begin{figure*}
\centering
\includegraphics[width=15cm]{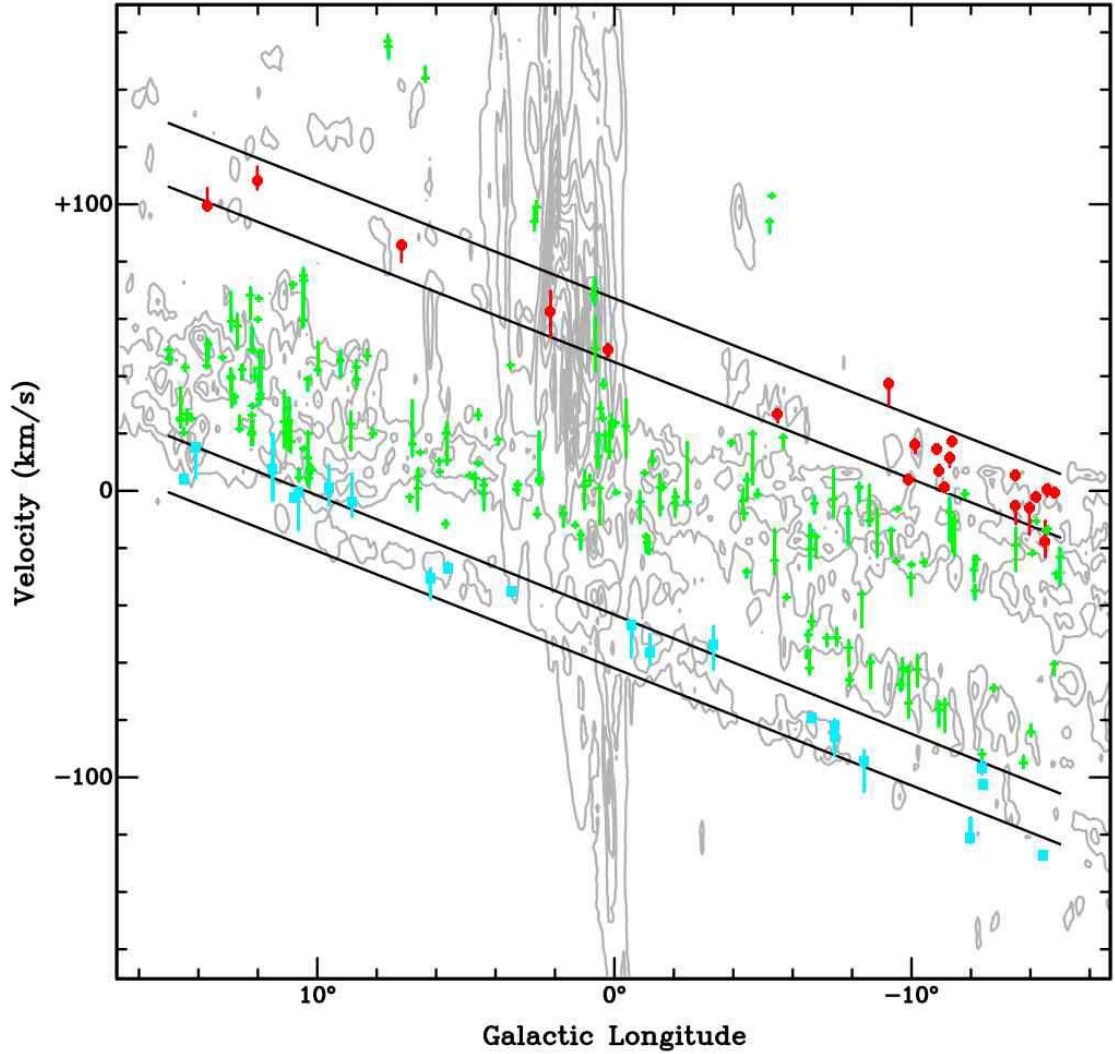}
\caption{\small 
Longitude-velocity diagram of 6.7-GHz methanol maser emission (symbols) 
and the CO 1$-$0 emission (contours) of \citet{dame08}. The CO 1$-$0 emission is shown for a latitude of 0$^{\circ}$ with contours at 10 to 100 \% of the peak emission.
Black diagonal lines delineate the near and far 3-kpc arms as defined 
by \citet{dame08}. Symbols show the peak of the maser emission with the 
velocity extent of the line delineating the range of velocity over which 
emission is seen. The 21 blue squares show the masers located in
the same longitude-velocity space as the near 3-kpc arm, the 21 red circles
those located in the far 3-kpc arm. The green crosses show the 6.7-GHz methanol masers
not associated with the arms, including the three masers associated with 
Sagittarius B2 and the four masers with comparable velocities to the 3-kpc arms,
but large latitudes (see main text).
Beyond longitude  $+10^{\circ}$ for the
near arm (and $-10^{\circ}$ for the far arm) the 3-kpc arm velocity space overlaps with spiral arm models and there is less certainty of the association. 
}
\label{LVplott}
\end{figure*}

The region delineated by the 3-kpc arms was observed for 6.7-GHz methanol masers 
with the 64-m Parkes radio telescope as part of the Methanol Multibeam (MMB) Survey, 
the full techniques of which are detailed in \citet{green09}. The MMB has detected over 
200 6.7-GHz methanol masers in the region $-15^{\circ} < l < 15^{\circ}$ (Caswell et al. in prep), of which 
49 (23 new to the survey, 26 previously known) have the velocity of their peak flux density emission
matching that of the near and far 3-kpc arms outlined by \citet{dame08}. These 49 masers 
are shown, together with their velocity ranges of emission, in  Fig.\,\ref{LVplott}. Three masers towards the
Galactic centre are most likely associated with Sagittarius B2, rather than the far 3-kpc arm (G000.650$-$0.067, G000.666$-$0.050 and G0.700$-$0.050). The CO emission of the near and far arms have Full Widths at Half Maximums (FWHMs) of 1.1$^{\circ}$ and 0.52$^{\circ}$ respectively \citep{dame08}. Assuming full widths to zero of approximately 2.2$^{\circ}$ and 1.04$^{\circ}$ means we can exclude any sources with latitudes outside this range. Four sources, G011.500$-$1.484, G345.417$-$0.950, G345.500$+$1.467 and G348.200$+$0.767, are excluded by this criterion (the first coincides with the near-side Norma arm and the middle two the near-side Carina-Sagittarius arm). The remaining 42 masers have peak velocities coincident with the CO emission velocities of the 3-kpc arm features (Fig.\,\ref{LVplott}) and show a clear association with the longitude-latitude distribution of the velocity integrated CO emission (Fig.\,\ref{LBPlott}), providing strong evidence for the existence of high-mass star formation in these regions.

The near and far side 3-kpc arms have distances differing by approximately a factor of two \citep{dame08}, 
which causes the latitude distributions in CO emission to also differ by a factor of two. We might therefore expect the masers
associated with the arms to also show this difference. Likewise, we might expect the flux densities to differ on average by a factor of four. 
Whilst we do not see these behaviours conclusively, we do see suggestions.  The 3-kpc arm masers, 21 in 
the near arm, 21 in the far arm, have a mean latitude of $-$0.050$^{\circ}$ 
with a standard deviation of 0.216$^{\circ}$. This compares with the mean latitude of all the MMB sources in the region, which is $-$0.102$^{\circ}$ 
with a standard deviation of 0.367$^{\circ}$. A Kolmogorov-Smirnov (KS) 
test finds no statistically significant evidence for a difference between the latitude distribution in the 3-kpc arm 
sample and that of the complete sample of masers in the longitude range. The individual arm samples do however
show a difference in distributions 
(Fig.\,\ref{LatPlot}), with the near having a mean latitude of $-$0.091$^{\circ}$ with a standard
deviation of 0.255$^{\circ}$ and the far having a mean latitude of 0.040$^{\circ}$ and a standard deviation
of 0.137$^{\circ}$. A KS test shows a difference in the near and far arm distributions at a greater than 95\% confidence level. With regards to the flux density, we see a difference in the medians, but not by the expected factor of four, 
with the far side sources (excluding the Sagittarius B2 associations) having a lower median 
flux density of $\sim$2.3\,Jy compared to the near side sources with a higher median flux density of $\sim$4.4\,Jy. 
Unfortunately these  comparisons are limited by small number statistics, preventing detailed analysis of the distributions. This is exacerbated by potential biasing of the far 3-kpc arm sample by a larger proportion of ambiguous (see next section) sources, together with fewer detections of weak sources in the far arm.

\subsection{Spiral arm cross-over}
\begin{figure*}
\centering
\includegraphics[width=8cm]{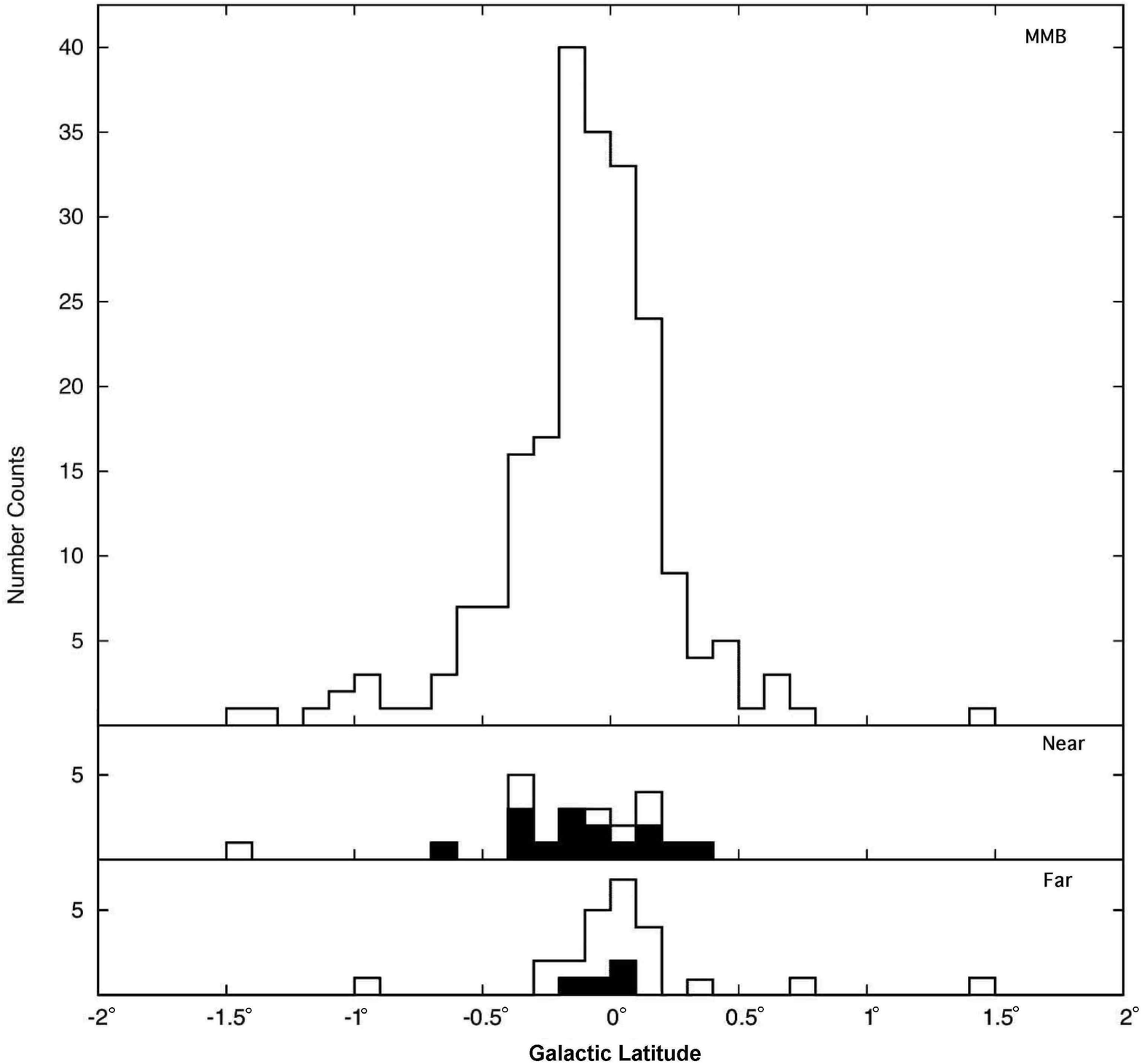}
\caption{\small Latitude distribution of 6.7-GHz methanol masers for the inner 15$^{\circ}$ of Galactic longitude.
Top: all the detections of the MMB. Middle: 22 sources associated with the near 3-kpc arm, with solid black showing the unambiguous associations (those that lie outside of the spiral arm crossover regions described in the text). Bottom: 24 sources associated with the far 3-kpc arm (excluding the 3 sources associated with Sgr B2), with solid black again showing the unambiguous associations. The four outlying sources (one in the middle plot and three in the bottom plot) are the four sources identified in the text as probably located nearer than the 3-kpc arms.
The latitude spread of the far arm (bottom plot) is clearly smaller than for the near arm (middle plot), a difference
mimicking that seen in CO.}
\label{LatPlot}
\end{figure*}

The kinematics of small regions of the far 3-kpc arm between $-15^{\circ} < l  < -10^{\circ}$ and the near 3-kpc arm between $+10^{\circ} < l < +15^{\circ}$ 
coincides with the spiral arm loci of the commonly adopted model of Galactic structure of  \citet{cordes02, cordes03}, when it is applied to a typical rotation
curve such as that of \citet{brand93}. The 
majority of this cross-over is with far-side spiral arms, but these are largely extrapolations of the logarithmic fits:  for example, the 
Carina-Sagittarius arm is not significantly traced by either CO or HI at longitude greater than 315$^{\circ}$ \citep[e.g.][respectively]{dame01, hartmann97}. However, it is apparent in the CO emission that more features exist 
in this region, and therefore some caution should be applied. Hence only those masers outside these cross-over regions are considered `unambiguous' (Fig.\,\ref{LatPlot}).

\section{Summary}
The 3-kpc arms have previously been believed to be devoid of significant star-formation, but the
current study shows not only star-formation, but high-mass star formation is present in the newly discovered
far 3-kpc arm and the well known near 3-kpc arm.  Although
the Galactic centre is a complex region in which there are a number of interpretations for the patterns 
and structures seen in longitude-velocity space, there is strong evidence that the 3-kpc arm features are 
real, and the 6.7-GHz methanol maser detections of the MMB survey 
have shown that it is very likely they exhibit high-mass star formation.
As a consequence these results imply high-mass star formation should be included in future models 
of the inner structure of our Galaxy.


\begin{figure*}
\centering
\includegraphics[width=15.1cm]{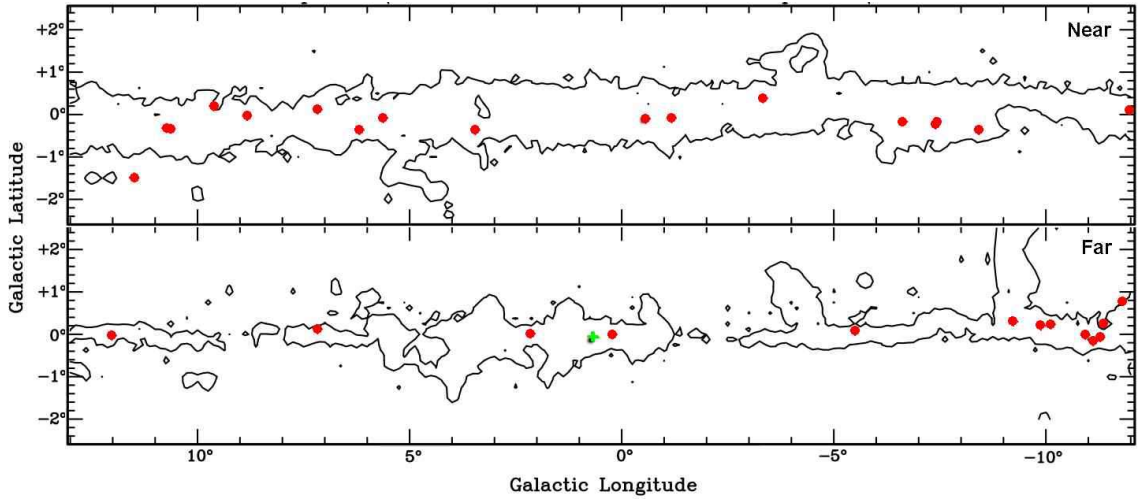}
\caption{\small Longitude-latitude distribution of the velocity integrated CO 1$-$0 emission of \citet{dame08} showing the boundary of the emission above 5 K kms$^{-1}$. The emission is integrated over velocities where the 3-kpc arms 
are believed to exist (see \citet{dame08} for details). Overlaid are the 6.7-GHz methanol masers associated with the 3-kpc arms (red circles) and the Sgr B2 sources (green crosses) from Fig.\,\ref{LVplott}.
The masers lie within the CO emission boundaries with the exception of G011.500$-$1.484 in the near arm and G348.200$+$0.767 in the far arm, however we suggest in the text that these should be excluded from the 3-kpc arm population.}
\label{LBPlott}
\end{figure*}

\acknowledgments
The authors thank T. Dame for providing the velocity integrated CO data used in Fig.\,\ref{LBPlott} and the referee for insightful comments. The Parkes Observatory is part of the Australia Telescope which is funded by the
Commonwealth of Australia for operation as a National Facility managed by
CSIRO. 

{\it Facilities:} \facility{Parkes (Methanol Multibeam)}.

\clearpage


\begin{thebibliography}{}
\bibitem[\protect\citeauthoryear{{Brand} \& {Blitz}}{{Brand} \&  {Blitz}}{1993}]{brand93} {Brand} J.,  {Blitz} L.,  1993, A\&A, 275, 67
\bibitem[Busfield et~al.(2006)]{busfield06} Busfield, A.~L., Purcell, C.~R., Hoare, M.~G., Lumsden, S.~L., Moore,  T.~J.~T., 
Oudmaijer, R.~D.,  2006, \mnras, 366, 1096
\bibitem[Caswell \& Haynes(1987)]{caswell87b} Caswell, J.~L., Haynes, R.~F., 1987, \aap, 171, 261
\bibitem[Caswell(1996)]{caswell96c} Caswell, J.~L.,  1996, MNRAS, 283, 606
\bibitem[Cersosimo(1990)]{cersosimo90} Cersosimo, J.~C.,  1990, \apj, 356, 156
\bibitem[Cordes \& Lazio(2002)]{cordes02} Cordes, J.~M., Lazio, T.~J.~W.,  2002, ArXiv Astrophysics e-prints, 0207156
\bibitem[Cordes \& Lazio(2003)]{cordes03} Cordes, J.~M.,  Lazio, T.~J.~W.,  2003, ArXiv Astrophysics e-prints, 0301598
\bibitem[\protect\citeauthoryear{{Cragg}, {Sobolev} \& {Godfrey}}{{Cragg}
  et~al.}{2005}]{cragg05}
{Cragg} D.~M.,  {Sobolev} A.~M.,    {Godfrey} P.~D.,  2005, MNRAS, 360, 533
\bibitem[Cyganowski et al.(2008)]{cyga08} {Cyganowski}, C.~J., et al., 2008, \aj, 136, 2391
\bibitem[\protect\citeauthoryear{{Dame}, {Hartmann} \& {Thaddeus}}{{Dame}  et~al.}{2001}]{dame01} {Dame} T.~M.,  {Hartmann} D.,    {Thaddeus} P.,  2001, ApJ, 547, 792
\bibitem[Dame \& Thaddeus(2008)]{dame08} Dame, T.M., Thaddeus, P., 2008, \apjl, 683, 143  
\bibitem[\protect\citeauthoryear{{Ellingsen}}{{Ellingsen}}{2006}]{ellingsen06}
{Ellingsen} S.~P.,  2006, \apj, 638, 241
\bibitem[Green et al.(2009)]{green09} Green, J. A., et al., 2009, \mnras, 392, 783
\bibitem[\protect\citeauthoryear{{Hill}, {Burton}, {Minier}, {Thompson},
  {Walsh}, {Hunt-Cunningham} \& {Garay}}{{Hill} et~al.}{2005}]{hill05}
{Hill} T.,  {Burton} M.~G.,  {Minier} V.,  {Thompson} M.~A.,  {Walsh} A.~J.,
  {Hunt-Cunningham} M.,    {Garay} G.,  2005, \mnras, 363, 405
\bibitem[Hartmann \& Burton(1997)]{hartmann97} Hartmann, D., Burton, W.B., 1997, Atlas of Galactic Neutral Hydrogen, Cambridge University press, Cambridge, UK
\bibitem[Lin \& Shu(1964)]{lin64} Lin, C.C., Shu, F.H., 1964, \apj, 140, 646
\bibitem[Lockman(1980)]{lockman80} Lockman, F.J., 1980, \apj, 241, 200
\bibitem[Menten(1991)]{menten91} {Menten} K.~M.,  1991, \apj, 380, 75
\bibitem[Minier et al.(2003)]{minier03} Minier, V., Ellingsen, S.~P., Norris, R.~P., Booth, R.~S., 2003, \aap, 403, 1095
\bibitem[\protect\citeauthoryear{{Minier}, {Burton}, {Hill}, {Pestalozzi},
  {Purcell}, {Garay}, {Walsh} \& {Longmore}}{{Minier} et~al.}{2005}]{minier05}
{Minier} V.,  {Burton} M.~G.,  {Hill} T.,  {Pestalozzi} M.~R.,  {Purcell}
  C.~R.,  {Garay} G.,  {Walsh} A.~J.,    {Longmore} S.,  2005, A\&A, 429, 945
 \bibitem[\protect\citeauthoryear{{Purcell}, {Balasubramanyam}, {Burton} \& {et
  al.}}{{Purcell} et~al.}{2006}]{purcell06}
{Purcell} C.~R.,  {Balasubramanyam} R.,  {Burton} M.~G.,    {et al.} 2006,
  MNRAS, 367, 553
\bibitem[Roberts(1969)]{roberts69} Roberts, W.W., 1969, \apj, 158, 123 
\bibitem[van Woerden et al.(1957)]{vanwoerden57} van Woerden, H., et al., 1957, CR Acad. Sci. Paris, 244, 1691
\bibitem[van der Walt et al.(2007)]{walt07} van der walt, D.J., Sobolev, A.M., Butner, H., 2007, \aap, 464, 1015
\bibitem[Xu et al.(2008)]{xu08} Xu, Y., Li, J.J., Hachisuka, K., Pandian, J.D., Menten, K.M., Henkel, C., 2008, \aap, 485, 729
\end{thebibliography}
\end{document}